\documentclass[apjl]{emulateapj} 
\usepackage{color}

\def\teff{\mbox{T$_{\rm eff}$}}
\def\logg{\mbox{log~{\it g}}}
\def\vmicro{\mbox{$\xi_{\rm t}$}}

\shorttitle{Sodium and Oxygen abundances in HB stars in NGC6121}
\shortauthors{Marino et al.}
\usepackage{ulem}

\begin{document}

\title{Sodium-Oxygen anticorrelation among Horizontal Branch Stars in the
  Globular Cluster M4 
          \footnote{Based  on data collected at the European Southern
    Observatory with the VLT-UT2, Paranal, Chile.       }}

\author{
   A.\ F.\ Marino\altaffilmark{1},
	  S.\ Villanova
	  \altaffilmark{2},
          A.\ P.\ Milone
	  \altaffilmark{3},
	  G.\ Piotto
          \altaffilmark{3},
	  K.\ Lind
          \altaffilmark{1},
	  D.\ Geisler
	  \altaffilmark{2},
          and
	  P.\ B.\ Stetson
	  \altaffilmark{4}
}

\altaffiltext{1}{Max-Planck-Institut f{\"u}r Astrophysik, Postfach 1317,
  D--85741 Garching b. M{\"u}nchen, Germany; amarino@mpa-garching.mpg.de} 

\altaffiltext{2}{Departamento de Astronom\'ia, Universidad de Concepci\'on,
  Casilla 160-C, Concepci\'on, Chile } 

\altaffiltext{3}{Department of Astronomy, University of Padova, Vicolo
  dell'osservatorio 3, 35122, Padova, Italy} 

\altaffiltext{4}{Herzberg Institute of Astrophysics, National Research Council
  Canada, 5071 West Saanich Road, Victoria, BC V9E 2E7} 

\begin{abstract}
The horizontal branch (HB) morphology of globular
clusters (GC) is mainly governed by metallicity. The second
parameter problem, well known since the 60's,
states that metallicity alone is not enough to describe the observed
HB morphology of many GCs.
Despite many efforts to resolve this issue, the second parameter
phenomenon still remains without a satisfactory explanation.
We have analyzed blue, red-HB, and RR-Lyrae stars in the GC M4 and 
studied their Fe, Na, and O abundances. Our goal is to
investigate possible connections between the bimodal HB of M4 and the chemical
signatures of the two stellar populations recently discovered among red giants
of this cluster.  
We obtained FLAMES-UVES/GIRAFFE spectra of a sample of 22 stars
covering the HB from the red to the blue region.
While iron has the same abundance in both the red and blue-HB segment, the
red-HB is composed of stars with scaled-solar sodium abundances, while the
blue-HB stars are all sodium enhanced and oxygen depleted.  
The RR-Lyrae are Na-poor, as the red-HB stars, and O-rich.
This is what we expect if the blue-HB consists of a second
generation of stars formed from the ejecta produced by an earlier stellar
population through high-temperature hydrogen-burning processes that
include the CNO, NeNa, and MgAl cycles and are therefore expected to be He-rich.
According to this scenario, the sodium and oxygen pattern detected in
the blue and red-HB segments suggests helium as the second parameter that
rules the HB morphology in M4.

\end{abstract}


\section{Introduction}
\label{introduction}

It is widely accepted that metallicity is the main parameter governing
the horizontal branch (HB) morphology of globular clusters (GC), i.e.\
metal-rich GC stars preferentially populate the red side of the RR-Lyrae
instability strip while metal-poor GCs tend to have blue HBs (e.g.\ Arp et
al.\ 1952). 
However, since the 1960's several observations revealed
that some GCs with similar
metallicity  exhibit different HB morphologies suggesting
that a second parameter is needed to fully account for the HB extent
(Sandage \& Wallerstein 1960; Sandage \& Wildey 1967).

Despite many efforts to understand the nature of the
second parameter, it still remains one of the open issues of modern
astrophysics. Several solutions have been proposed, including age,
stellar mass, cluster central density, cluster mass
(see Dotter et al.\ 2010 and
references therein). Interestingly, a difference in helium could
explain the observed HB morphology in GCs
(van den Bergh 1967).

In this context, the recent discovery of multiple stellar populations
in GCs (Piotto et al.\ 2007; Milone et al.\ 2008, 2010; Marino et al.\ 2009)
allows us to look at the second parameter problem from a new point of view, as
the multiple HB components of several GCs have been tentatively assigned 
to different stellar generations by many authors
(e.g.\ D'Antona et al.\ 2005, Milone et al.\ 2008).

The aim of this paper is to test whether stellar populations may be
linked to the HB morphology in the nearby GC M4 (=NGC~6121).
This cluster is an ideal target, as both spectroscopic and photometric
studies have provided evidence that it experienced multiple star-formation
events. 

Chemical composition analysis of bright red giant branch (RGB) stars showed
star-to-star variations in the C, N, O, Na, and Al abundances
(Gratton et al.\ 1986, Brown et al.\ 1990, Brown \& Wallerstein 1992, Drake et
al.\ 1992), that define a clear CN distribution 
bimodality (Norris 1981), a Na-O anticorrelation,
and a Na-Al correlation (Ivans et al.\ 1999; Marino et al.\ 2008, hereafter
M08).  

A bimodality in the distribution of stars on the Na-O anticorrelation in M4
has been found the recent analysis of high resolution spectra of 105 RGB stars
by M08, who identified two stellar populations with a 
 strong dichotomy in Na and O abundance. Na-rich stars
are also CN-strong and are more Al-enriched than Na-poor
ones. Neither any Mg-Al anticorrelation nor significant star-to-star
variations in iron, iron-peak elements, {\it s}-process,
or $\alpha$ elements were detected.

The presence of two stellar populations in M4 is confirmed by
photometry. The two groups of Na-rich/O-poor and Na-poor/O-rich stars define
two distinct sequences in the $U$-$(U-B)$ color-magnitude
diagram (CMD).
Na-rich stars populate a sequence on the red side of the RGB, while
Na-poor ones define a bluer, broader sequence.

Interestingly, M4 has a bimodal HB, well populated both on the
blue and the red side of the RR-Lyrae gap. 
The existence of a physical connection between the CN-dichotomy observed
on the RGB of this cluster and the HB morphology
(the so-called second parameter problem) was already suggested 
by Norris (1981). 
The first evidence of such a connection came from Smith \& Norris (1993), who
found that the majority of the red-HB stars in M4 have a similar strong CN
content. 
Nevertheless, from the Smith \& Norris (1993) study one could tentatively
associate the red-HB with the N-rich/C-poor/Na-rich stars, the analysis of one
blue-HB star done by Lambert et al.\ (1992), reveals that it could be the
counterpart of a C-poor RGB star. 
We note, however, that Smith \& Norris (1993) expressed caution concerning
their absolute measures of the CN abundances, but stated that the CN abundance
was similar in the analyzed red-HB stars. 

In this paper we investigate the possibility that the HB bimodality in M4
could be related with the two stellar populations identified by M08  
by investigating chemical abundances directly on HB stars.

\section{Observations and data analysis}
\label{data}
The data set consists of high and medium resolution spectra obtained with the
FLAMES/UVES (set-up 520nm, R$\sim$47,000) and FLAMES/GIRAFFE (set-ups HR11,
HR13, HR15N, R$\sim$20,000) spectrographs (runs: 71.D-0205A, 083.B-0083A). 
These data were reduced using the dedicated pipelines ({\sf
  http://girbld-rs.sourceforge.net} for GIRAFFE and Ballester et al.\ 2000 
  for UVES).
On the basis of the position on the CMD, represented in Fig.1 (see
  Sect~2.1 for more details), our sample includes 6 blue-HB (blue symbols) and
  16 stars distributed on the red side (red symbols). Two of the latters 
have been observed with UVES (red circles) and 14 with GIRAFFE (red crosses). 
For 16 red stars we derived Fe and Na, and, for the UVES targets only, O
abundances.

Each star observed with UVES has from 2 to 4 exposures, and the final combined
spectra have a typical S/N ranging from $\sim$60 in the bluer part of the
spectrum, to $\sim$90 in the redder one ($\lambda$$\sim$6150 \AA), where
  the O triplet that we measured is located. 
GIRAFFE spectra for each star were obtained by co-adding 10 exposures,
  and have a higher S/N, ranging from $\sim$150 for the HR11 set-up to
  $\sim$250 for the HR15N one.

We have identified all but one (GIRAFFE \#47328), stars in the catalog of
  Cudworth \& Rees (1990), and found that the two stars observed with UVES on
  the red part of the HB are RR-Lyrae variables listed in the catalog by
  Sawyer-Hogg (1973).

Chemical abundances for iron were derived from equivalent widths (EW), by
  using the local thermodynamical equilibrium (LTE) code MOOG (Sneden 1973),
  coupled with model atmospheres interpolated from the grid of ATLAS9 models
  (Castelli \& Kurucz 2004). The linelist was extracted by M08, and the atomic
  linelist by Qiu et al.\ (2001), more appropriate for warmer stars. We used
  Fe lines to derive the atmospheric parameters (\teff, \logg, \vmicro, and
  [Fe/H]\footnote{We use [Fe/H] synonymously with stellar overall
    metallicity.}): \teff\ and \vmicro\ were
    derived by removing trends in  \ion{Fe}{1} abundances with excitation
    potential and with EWs respectively; gravities \logg\ by satisfying the
    ionization equilibrium between \ion{Fe}{1} and \ion{Fe}{2}
    abundances. This process was done iteratively until a final interpolated
    model was obtained.
  
Sodium was measured from EWs of the doublets at $\sim$5890 \AA\ and
  $\sim$5685 \AA.\ For the blue-HB stars, that have \teff$\gtrsim$8000, we
  were not able to measure the weak doublet at $\sim$5685 \AA,\ while for the
  two RR-Lyrae observed with UVES, we derived Na from the weak doublet,
  discarding the very strong lines at $\sim$5890 \AA, since at these
  temperatures, they get too strong for accurate abundance measurements. For
  stars observed with GIRAFFE, due to the lower resolution, we used a spectral
  synthesis analysis of the weak doublet at $\sim$5685 \AA.\ As Na spectral
  lines for our stars are heavily affected by departures from LTE that depends
  also on the spectral line strengths, we derive sodium in the non-LTE regime,
  by generating non-LTE profiles using the non-LTE spectral code MULTI
  (Carlsson 1986; see Lind et al.\ subm. for a description of the non-LTE
  modeling) coupled with our Kurucz model atmospheres. As an example, we show
  in the upper panel of Fig.~3 the observed Na lines for the blue-HB star
  \#22746 with superimposed the non-LTE profiles. We quoted the Na abundances
  obtained for this star in the non-LTE regime. 

Oxygen was estimated from spectral synthesis to the triplet at $\sim$6157
\AA\ available for UVES. Synthetic spectra at different O were matched
  with the observed spectrum, and we choose the best fit model by minimizing
  the $\chi^{2}$ with respect to the observed spectrum. For O, expected
non-LTE effects ($\lesssim$0.1 dex) are less significant than for Na, hence we
applied corrections available in the literature (Takeda et al.\ 1997).   

Non-LTE effects could also affect the Fe lines. However, in our analysis, we
did not consider non-LTE corrections for Fe lines, that should not affect
significantly the  ionization equilibrium (Villanova et al. 2009). 

To investigate the reliability of our atmospheric parameters we did the tests
represented in Fig.~2. In the left panel, we compared our adopted \teff\ with
those derived by isochrones by D'Antona et al.\ (2009), coupled with our
photometric data. Our sample of stars distributes with a dispersion of
$\sim$50 K around the line of perfect agreement. The dispersion is similar for
the red-HB stars observed with GIRAFFE and the blue-HB stars observed with
UVES. This value has been taken as a rough estimate of the error associated to
our temperatures, of course this is only internal. As expected since the
photometric data have not been taken contemporary to the spectra, the two
RR-Lyrae show a larger scatter from the line of perfect agreement. To verify
wheater our \teff\ scale is correct, we compared the observed $(B-V)$ colors
with the synthetic Kurucz's colors $(B-V)$$_{\rm Castelli}$ (Castelli
1999). For this purpose we dereddened our colors by assuming E$(B-V)$=0.36
(Harris 1996). These tests are shown in the two right panels of
Fig.\ref{errori}, where the observed $(B-V)$ and the adopted \teff\ for the
blue-HB stars (upper panel), and red-HB plus RR-Lyare stars (lower panel) have
been compared with the $(B-V)$$_{\rm Castelli}$ at [A/H]=$-$1, for different
gravities. Figure~2 suggests that the internal uncertainty associated to our
\logg\ measurements is 0.25. Red-HB stars distribute in agreement with what
expected from synthetic colors. Their position on the \teff-$(B-V)$$_{\rm
  Castelli}$  plane are much less sensitive to gravity (as an example we
plotted the tracks corresponding to \logg=2.5 and \logg=4.0). However, since
these stars have almost the same magnitude, we expect them to have a similar
gravity. Hence, the rms of the \logg\ values obtained for these stars, that is
0.17, could be taken as an estimate for the internal uncertainty related to
gravities. The error in \vmicro\ has been estimated following the procedure
described in M08: i) we calculated for each star the error associated with the
slope of the best least squares fit in the relation between Fe~I abundances
vs.\ reduced EWs; ii) we varied microturbolece until the slope of the 
  best fitting line in the Fe~I-EWs relation is equal to this error;
iii) the corresponding difference in the \vmicro,\ ($\sim$0.20 km/s), has been
taken as an estimate of the internal uncertainty associated with this
parameter.   

The limited S/N of the spectra imply an error in the placement of the
continuum both for the EW measurements and the spectral synthesis. The mean
difference in EWs among the same lines observed in pairs of stars with similar
atmospheric parameters, that is $< \Delta$EW$> \simeq$5 m\AA\ for UVES, and
slightly lower for GIRAFFE, thanks to the high S/N, is an estimate of these
uncertainties. To verify how model atmosphere and EW uncertainties influence
the derived chemical abundances, we repeated the abundance measurements for
one RR-Lyrae star (\#34211), one blue-HB star (\#27613), and one red-HB star
(\#46113). For this exercise, one by one at a time, we changed the uncertain
parameters by $\Delta$\teff=$\pm$50 K, $\Delta$\logg=$\pm$0.25,
$\Delta$\vmicro=$\pm$0.20,  $\Delta$[A/H]=$\pm$0.2, $\Delta$EW=5~m\AA
respectively.  
Assuming that these uncertainties are uncorrelated, we estimate total
abundance uncertainties by summing in quadrature the various 
contributions. The resulting internal errors are typically of $\sim$0.10 dex
for the iron abundances, and $\sim$0.05 dex for the Na abundance ratios
relative to iron. We also test the [Na/Fe] sensitivities to changes in
temperature of $\pm$100 K, and verified that the Na abundance ratios are
affected at a level of $\lesssim$0.10. Oxygen abundance ratios are marginally
affected by model uncertainties: while its abundance ratio is almost
insensible to changes in \teff,\ changes in 
\logg\ of $\pm$0.2 affect the  [O/Fe] abundances of $\mp$0.02. 
However, in this case, 
 the dominant error source
is the continuum placement, that strongly depends on the S/N of our spectra,
and allows us to define the best fit among the observed and synthetic spectra
with an uncertainty of $\sim$0.10 dex.  
An example of the O synthesis has been shown  in the lower panel of
Fig.~3, where we show the best fit synthesis, and
deviations of $\pm$0.2 from the best fit.

The linelist with derived EWs, is listed in the on-line available Table~2.
The photometry, adopted atmospheric parameters, chemical contents, and applied
non-LTE corrections, are listed in Table~1. 

\subsection{Photometry}
Our photometry provided by Stetson, has been already described in D'Antona et
al.\ (2009). 
The observed colors and magnitudes have been corrected for differential
reddening, and proper motions (Anderson et al.\ 2006) are used to
separate field from cluster stars.

In the $V$-$(B-I)$ CMD 
corrected for differential reddening 
of Fig.~1 we plotted only stars assumed to be cluster members, separated from
the field as shown in the right-hand panels, with the 
proper motion vector point diagrams for four magnitude intervals. In
  these panels DX and DY are the proper motions in pixels. 
Proper motions are calculated with respect to a sample of cluster stars, and
therefore member stars define a narrow bulk of stars centered around DX=DY=0. 
Red circles isolate the objects that have member-like motions (black dots).
To correct for differential reddening, we used the method described in Milone
et al.\ (2010, 2010 in prep). Briefly, we define the fiducial main sequence
and estimate for each star, how the stars in its vicinity may systematically
lie to the red or the blue of the fiducial sequence; this systematic color 
offset is indicative of the local differential reddening.

\section{Results}
\label{results}

In this section we present the obtained abundances, and compare chemical
properties of the blue, red-HB and RR-Lyrae stars.

Our analysis confirms that M4 has an average iron abundance of:
\begin{center}
$[\rm Fe/H]=-1.12\pm0.02$
\end{center}
(internal error) in agreement, within errors, with results obtained on RGB
stars by M08. 
This in turn confirms that our study is not affected by significant systematic
errors.  
There is no significant difference in the average [Fe/H] of stars located on
different HB segments, as the mean iron abundances derived for blue-HB stars
agrees, within the errors, with the one obtained for red-HB and RR-Lyrae
($[\rm Fe/H]_{\rm blue-HB}=-1.07\pm0.02$ and $[\rm Fe/H]_{\rm
  red-HB+RR-Lyrae}=-1.14\pm0.03$). 

On the contrary, we detected large star-to-star variations in [Na/Fe].
All the red-HB stars have lower sodium,
with [Na/Fe] ranging from $\sim-$0.2 to $\sim$+0.2 dex, average:

\begin{center}
$[\rm Na/Fe]_{\rm red-HB+RR-Lyrae}=-0.01\pm0.01$ 
\end{center}
(internal error), and observed dispersion of $\sim$0.10 dex.

Conversely, the blue-HB stars all have high sodium, between $\sim$+0.25 and
$\sim$+0.40, with average:
\begin{center}
$[\rm Na/Fe]_{\rm blue-HB}=+0.34\pm0.02$
\end{center}
(internal error) and an observed dispersion of $\sim$0.06 dex.

The two RR-Lyrae have a mean $[\rm Na/Fe]_{\rm RR-Lyrae}=0.04$, in
  concert with the red-HB stars. 

For one blue-HB star (\#36648), we were able to measure Na from one line
  of the weak doublet, that gives enhanced Na abundance as obtained from the
  resonance doublet. This comparison allow to safely discard systematics
  introduced by the analysis of different spectral lines as the cause of the
  observed differences in Na. 

In Fig.~4 we highlight the relation between the position of 
stars along the HB $(B-I)$ and their sodium abundances:
a clear anticorrelation
between the observed $(B-I)$ and the sodium abundance is present,
with blue and red-HB+RR-Lyrae stars clustered around two distinct values
of color and [Na/Fe]. 

Oxygen has been measured for the stars observed with UVES, i.e.\ the whole
sample of blue-HB stars and the two RR-Lyrae.   
The blue-HB stars have an average oxygen of:

\begin{center}
$[\rm O/Fe]_{\rm blue-HB}=+0.27\pm0.02$
\end{center}
(internal error only) and an observed dispersion of $\sim$0.05 dex.
The two variables for which we were able to see the O lines
have higher O content.
For GIRAFFE targets we could
not estimate the O content because the forbidden line is too weak.
The RR-Lyrae turn out to be enriched in O with a mean of $[\rm O/Fe]_{\rm
  red-HB}=+0.50\pm0.01$. 

The present O and Na results are in good agreement with those found
from high resolution UVES spectra of 105 RGB stars by
M08. In Fig.~4 we show the Na-O
anticorrelation for the RGB stars (gray squares) from M08, and the HB stars
analyzed here. We arbitrarily plotted the RHB stars observed with
GIRAFFE, for which oxygen abundances are not available, at [O/Fe]=+0.70.
As already discussed in Sect.~\ref{introduction}, from their study of
RGB stars M08 detected two stellar
populations in M4 with different sodium, oxygen, and CN abundances.
The two groups of Na-poor/O-rich and Na-rich/O-poor stars define
the Na-O anticorrelation of 
Fig.~4, and trace two branches on the RGB when the $(U-B)$ color is
  used. 

Apart from possible small zero-point displacements, likely due to the use of
different spectral lines and non-LTE corrections, the HB stars
analyzed here follow the trend defined by the giants.
The blue-HB stars lie in the O-poor/Na-rich group, while the red-HB ones
belong to the
O-rich/Na-poor stellar population.
Possible enhancements in He do not affect the [X/Fe] abundance
  measurements of these gravity-insensitive species (Lind et al.\ subm.). 

Interestingly, the two RR-Lyrae belong to the Na-poor/O-rich population,
  corresponding to the first stellar generation.  

\section{Conclusions}
\label{discussion}
In this paper we studied a sample of HB stars in the GC
M4, with the main purpose of measuring their abundance of sodium
and oxygen.
Our targets exhibit a bimodal Na distribution very similar to that
found for RGB stars, with red-HB stars having roughly solar-scaled [Na/Fe] and
blue-HB stars being all sodium enhanced. Blue-HB stars  are also oxygen
depleted, while the two variables are oxygen rich, and sodium poor as well as
the red-HB stars. 

The Na-O anticorrelation of HB stars is almost the same as we found
for RGB stars, where we detected two distinct stellar populations with
different mean Na and O.
{\it On the basis of their position in the Na-O plane, the blue-HB stars can
  be clearly associated with the second (O-poor/Na-rich) stellar
  generation,while the red-HB and the RR-Lyrae must be the progeny of the
  O-rich/Na-poor first generation.}  

As already mentioned in Sect.~\ref{introduction},
the Na-O anticorrelation is a powerful tracer of the star formation
history in a GC.
It has been observed among unevolved stars in GCs (Gratton et al.\ 2001),
demonstrating that this pattern is 
not the result of some mixing process but is rather
produced by the ejecta of a first stellar generation.
It is not a property
of a few peculiar objects but has been
observed in all the GCs studied so far.
While it is now widely accepted that it indicates the presence of
material gone through
high temperature H-burning processes, which include the CNO, NeNa,
and MgAl cycles, the nature of the polluters is still unclear.
These processes might have occurred in intermediate-mass asymptotic giant
branch stars (D'Antona et al.\ 2002) or in fast-rotating massive
stars (Decressin et al.\ 2007).
In any case, helium is the main product of H-burning and is
expected to be related to Na and O abundance, with stars that are
Na-enhanced and O-depleted being also He-enriched.

According to this scenario, the present day He-rich (Na-rich,
O-poor) stars should be less massive than He-poor (Na-poor, O rich)
ones because the latter evolve more slowly. As a consequence of this,
the blue-HB should consist of He/Na-rich, O-poor stars, while the He/Na-poor,
O-rich stellar population should end up on the red-HB (D'Antona et al.\ 2002).
The abundances of sodium and oxygen in HB stars presented in this
paper strongly support this picture.

Our results show that in M4 the spread of light elements
is strongly correlated with the HB morphology. It is
tempting to affirm that, at least in this cluster, this spread is the {\it
  second parameter},  
with He possibly being the main driver of its HB morphology after Fe.

{\it Acknowledgements} 
We thank the referee whose suggestions have improved significantly the paper, 
M.\ Bergemann, L.\ Mashonkina, F.\ D'Antona for useful discussions.
AFM, APM, and GP acknowledge the support by MIUR under the PRIN2007
(prot.20075TP5K9).

{}

\begin{figure}[ht!]
\centering
\plotone{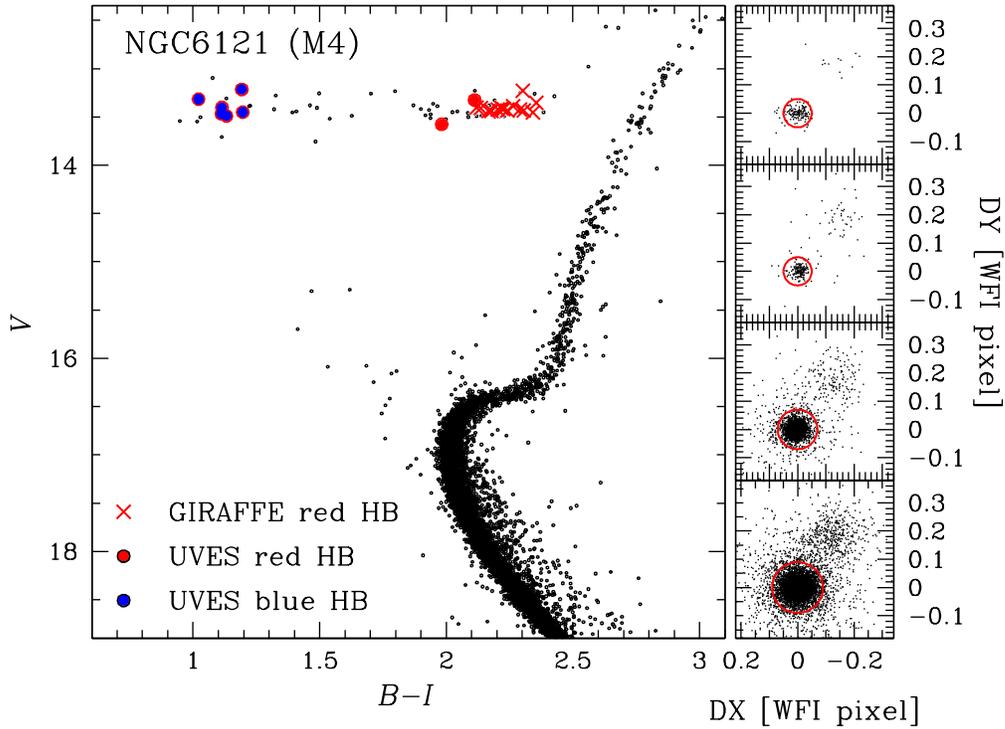}
\caption{CMD corrected for differential reddening for all stars
  assumed to be cluster members. UVES and GIRAFFE targets are marked with
  circles and crosses respectively. Blue and red colors 
  represent stars in the blue and red-HB, respectively.
On the right panels we represent
proper motion
 diagrams for stars in the M4 field of view in four
  magnitude intervals. A circle shows the adopted membership
  criterion.
         }
\label{CMD}
\end{figure}

\begin{figure}[ht!]
\centering
\plotone{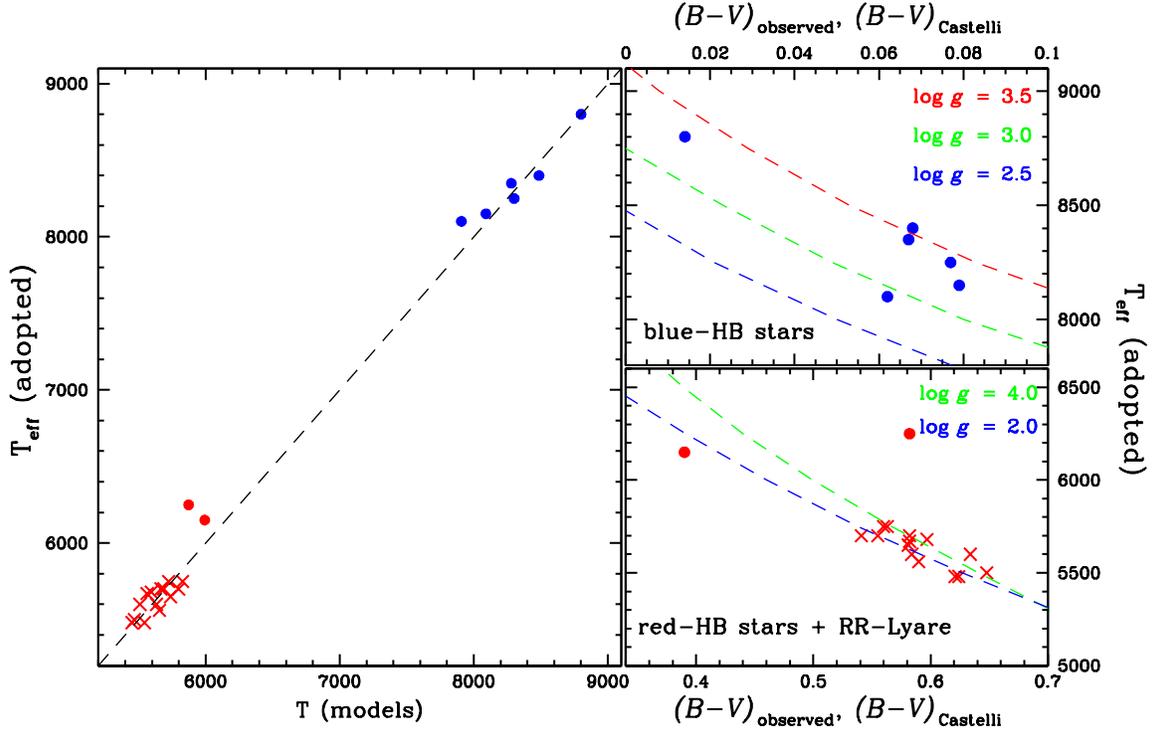}
\caption{$Left:$ Adopted \teff\ vs.\ of the temperature derived from
  isochrones and observed colors. Dashed line represents the perfect
  agreement. $Right:$ Adopted \teff\ versus observed ($(B-V)$$_{\rm
    observed}$)  for the blue-HB stars (upper punel) and the red-HB+RR-Lyrae
stars (lower panel). Dashed lines represent tracks obtained from the synthetic
colors ($(B-V)$$_{\rm Castelli}$) at different gravities. The
\logg\ corresponding to  each track has been quoted in both panels.      }

\label{errori}
\end{figure}

\begin{figure}[ht!]
\centering
\plotone{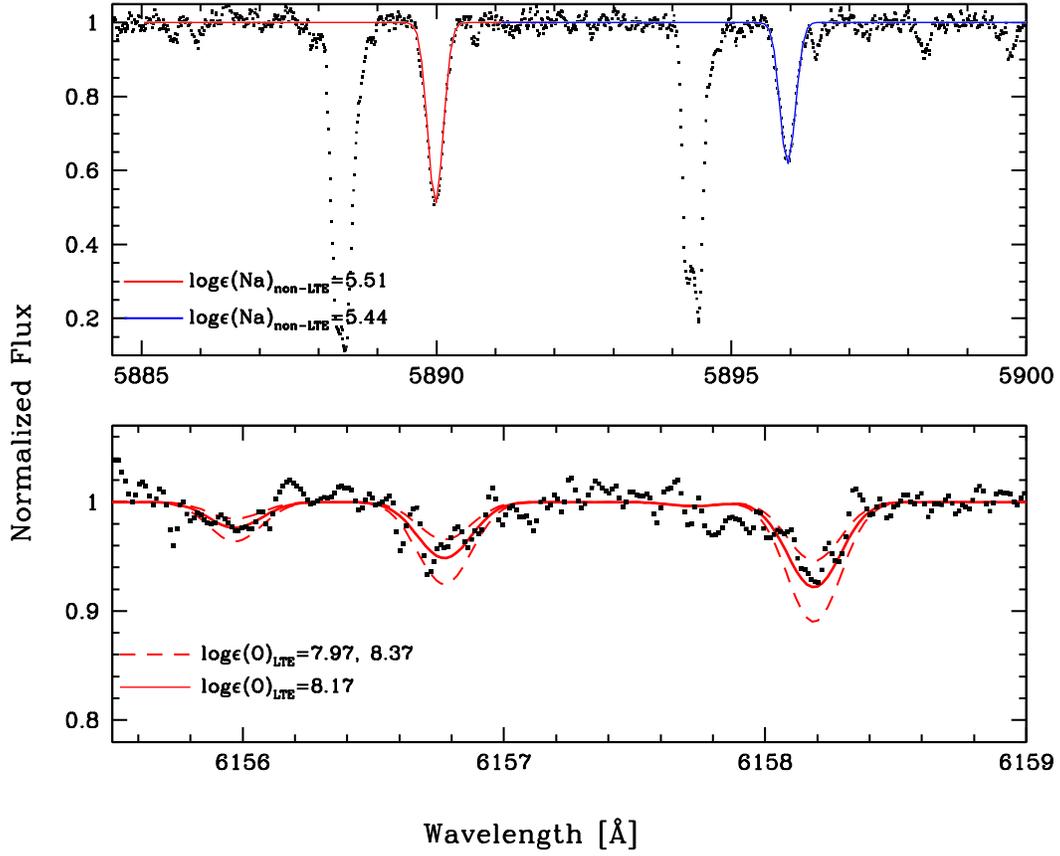}
\caption{
{\it Upper Panel:} non-LTE profiles for the two Na lines of the star
\#22746. The abundances obtained by each line is quoted in the panel.  
{\it Lower panel:} Example of O spectral line fitting for the star
\#36648. Synthetic spectra with differences in abundance of $-$0.20, $+$0.00,
$+$0.20 dex with respect to the adopted value are shown. 
}
\label{sintesi}
\end{figure}

\begin{figure}[ht!]
\centering
\plotone{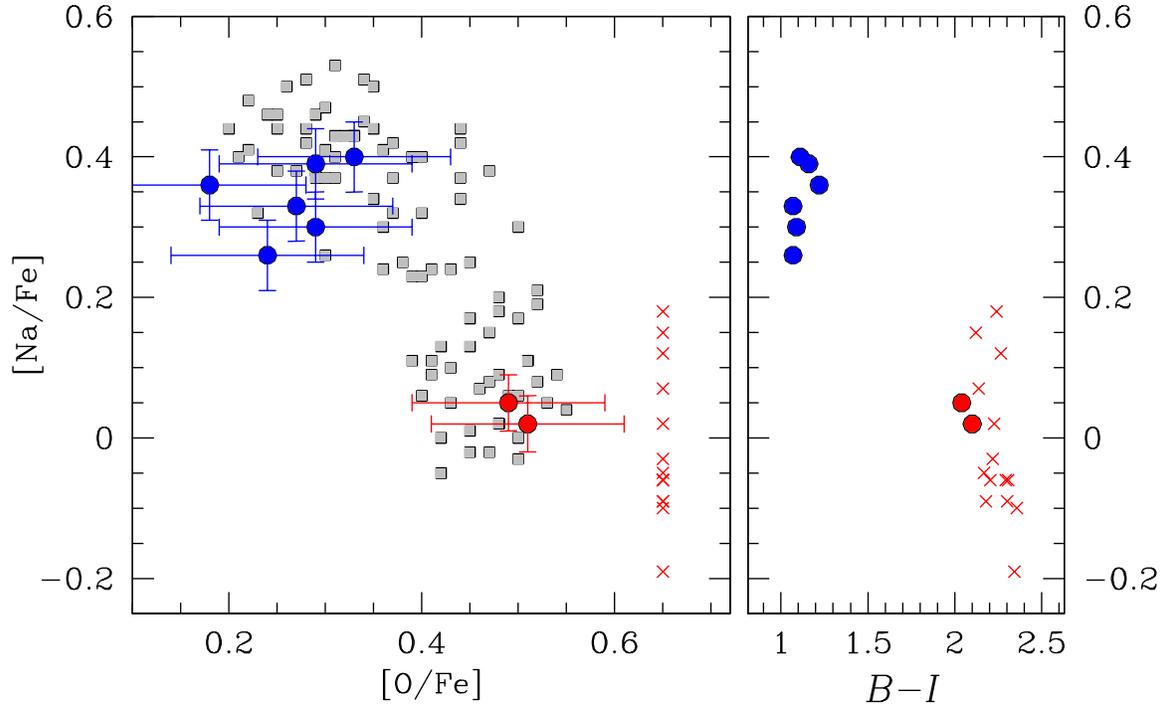}
\caption{
{\it Left}: [Na/Fe] versus [O/Fe] for UVES stars represented with blue and red circles. Red crosses indicate the GIRAFFE red-HB stars for which [O/Fe] is not
  available. As a comparison, we show as gray squares [Na/Fe] and 
  [O/Fe] for RGB stars from M08. {\it Right}: [Na/Fe] as a function of $(B-I)$.
}
\label{NavsBI}
\end{figure}

\clearpage

\end{document}